\documentclass[10pt]{article}
\usepackage{fancyhdr}
\usepackage{extramarks}
\usepackage{amsmath}
\usepackage{amsthm}
\usepackage{amsfonts}
\usepackage{siunitx}
\usepackage{tikz}
\usepackage[plain]{algorithm}
\usepackage{algpseudocode}
\usepackage{multirow}
\usepackage{booktabs}
\usepackage{graphicx}
\usepackage{subfigure}
\usepackage[colorlinks,linkcolor=black,anchorcolor=black,citecolor=black,urlcolor=blue]{hyperref}
\usepackage{amsmath,bm}
\usepackage{booktabs}
\usepackage{mathtools}
\usepackage{amssymb}
\usepackage{caption}
\usepackage{capt-of}
\usepackage{mciteplus}
\usepackage{cite}
\usepackage{mathrsfs}
\usepackage[title,titletoc,toc]{appendix}
\usepackage{xr}
\usepackage{parskip}
\usetikzlibrary{automata,positioning}
\usepackage{wrapfig}
\renewcommand{\part}[1]{\textbf{\large Part \Alph{partCounter}}\stepcounter{partCounter}\\}

\begin{document}

\title{Review of  the mechanisms of SARS-CoV-2 evolution and transmission}
\author{Jiahui Chen$^1$,  Rui Wang$^{1 }$\footnote{The first two authors contribute equally.}, 
 and Guo-Wei Wei$^{1,2,3}$\footnote{
Corresponding author.		E-mail: weig@msu.edu} \\
$^1$ Department of Mathematics, \\
Michigan State University, MI 48824, USA.\\
$^2$ Department of Electrical and Computer Engineering,\\
Michigan State University, MI 48824, USA. \\
$^3$ Department of Biochemistry and Molecular Biology,\\
Michigan State University, MI 48824, USA. \\
} 

\maketitle

\begin{abstract}
 
The mechanism of severe acute respiratory syndrome coronavirus 2 (SARS-CoV-2) evolution and transmission is elusive and its understanding, a prerequisite to forecast emerging variants, is of paramount importance.  SARS-CoV-2 evolution is driven by the mechanisms at molecular and organism scales and regulated by the transmission pathways at the population scale. In this review, we show that infectivity-based natural selection was discovered as the mechanism for SARS-CoV-2 evolution and transmission in July 2020. In April 2021, we proved beyond all doubt that such a natural selection via infectivity-based transmission pathway remained the sole mechanism for SARS-CoV-2 evolution. However, we reveal that antibody-disruptive co-mutations  [Y449S, N501Y] on the spike protein receptor-binding domain (RBD) debuted as a new vaccine-resistant transmission pathway of viral evolution in highly vaccinated populations a few months ago. Over one year ago, we foresaw that mutations on RBD residues, 452 and 501, would  ``both have high chances to mutate into significantly more infectious COVID-19 strains''. Mutations on these residues underpin prevailing SARS-CoV-2 variants Alpha, Beta, Gamma, Delta,  Epsilon, Theta, Kappa, Lambda, and Mu at present and are expected to be vital to emerging variants in the future.  
We anticipate that viral evolution will combine RBD co-mutations at these two sites, creating future variants that are about ten  times more infectious than the original SARS-CoV-2. 
Additionally, two complementary transmission pathways of viral evolution, i.e.,  infectivity and vaccine resistance will prolong our battle with COVID-19 for years. We predict that  RBD co-mutation sets [A411S, L452R, T478K], [L452R, T478K, N501Y],    [L452R, T478K, E484K, N501Y],   [K417N, L452R, T478K], and [P384L, K417N, E484K, N501Y] will have a high chance to grow into dominating variants due to their high infectivity and/or strong ability to break through current vaccines, calling for the development of new vaccines and antibody therapies.

\end{abstract}

 
\newpage

\section{Introduction}
The coronavirus disease 2019 (COVID-19) pandemic caused by severe acute respiratory syndrome coronavirus 2 (SARS-CoV-2) has led to over 225 million confirmed cases and over 4.6 million fatalities. Researchers have been racing against the devastation of COVID-19 in the past 22 months. Although there were two earlier outbreaks of deadly pneumonia caused by $\beta$-coronaviruses: SARS-CoV (2002) and Middle East respiratory syndrome coronavirus (MERS-CoV) (2012) in the 21st century, SARS-CoV-2  astonished the unprepared scientific community. Currently, Google Scholar has accumulated  188,000  items under ``COVID-19'' and  139,000 items under ``SARS-CoV-2'' since 2020. Nonetheless, the scientific community still does not know enough about SARS-CoV-2 and COVID-19 22 months after the outbreak of the pandemic. One of the greatest challenges of our time is the understanding of the mechanisms of SARS-CoV-2 evolution and transmission \cite{casalino2021ai,clark2021sars,he2020molecular}. The importance of such an understanding cannot be overemphasized. It is a prerequisite to forecast the emerging SARS-CoV-2 variants, which, in turn, is an imperative for the design of the next generation of mutation-proof vaccines and monoclonal antibodies (mAbs).   

\begin{figure} 
	\centering
	\includegraphics[width=0.98\textwidth]{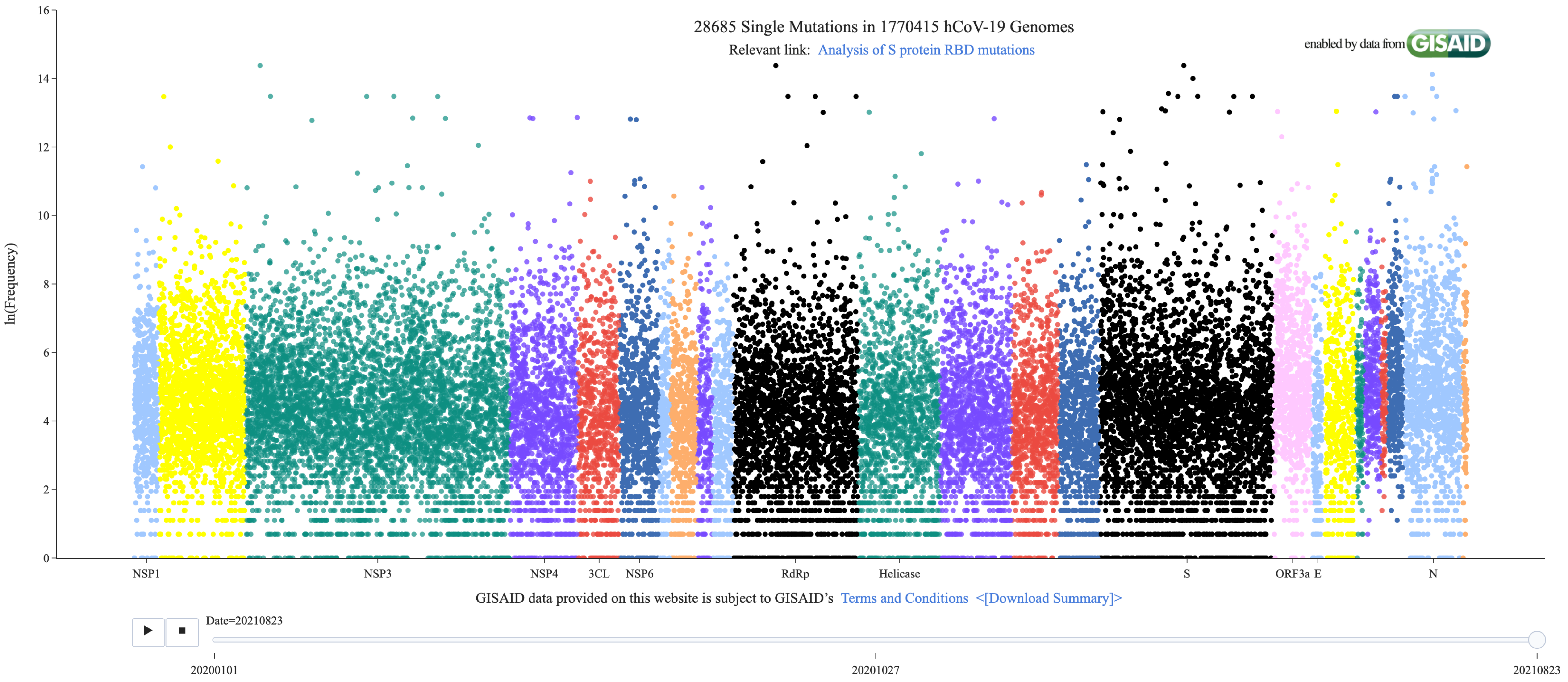}
    \caption{Illustration of  SARS-CoV-2 mutations given by  Mutation Tracker.  Interactive version is available at web site:  \url{ https://users.math.msu.edu/users/weig/SARS-CoV-2_Mutation_Tracker.html}. 
		}
    \label{fig:mutationTracker}
\end{figure}

  SARS-CoV-2 encodes 29 genes and its genome has about 29,900 nucleotides. As illustrated in our Mutation Tracker (see Figure \ref{fig:mutationTracker}), nearly 29,900 mutations are essentially evenly distributed on the whole viral genome, showing a confounding pattern. Additionally, the mutagenesis of SARS-CoV-2 genomes is driven by a large number of competing processes, including molecular-scale random shifts, replication errors, transcription errors, translation errors, proofreading, and recombination, organism scale host editing induced by the immune response and host-viral recombination, and population-scale natural selection. Moreover, discrepancies among reported experimental binding free energies can be over 100 fold for viral spike (protein) receptor-binding domain (RBD) complexes with angiotensin-converting enzyme 2 (ACE2) or antibodies  (see Table 1 of a Ref. \cite{chen2021review}), creating a baffling situation. These perplexing factors make the understanding of viral transmission and evolution one of the most challenging tasks. 
 
The recent global surge in COVID-19 infections has been fueled by new SARS-CoV-2 variants, namely Alpha, Beta, Gamma, Delta, Theta, Epsilon, Kappa, Lambda, Mu, etc. A common feature for these variants is that they all involve one of two spike (S) protein receptor-binding domain (RBD) residues 452 and 501. The importance of these sites were predicted by us more than one year ago. We foresaw that residues, 452 and 501, out of 194 RBD residues, “have a high chance to mutate into significantly more infectious COVID-19 strains”, by ``combining sequence alignment, probability analysis, and binding free energy calculation'' (see the last sentence in the Abstract of Ref.  \cite{chen2020mutations}). Additionally, on page 5218 of the aforementioned paper \cite{chen2020mutations}, we also predicted three lists of RBD mutations: 1149 most likely, 1912 likely, and 625 unlikely for a total of 3,686 possible RBD mutations. Currently, essentially all 724 observed RBD mutations belong to the list of most likely ones. Finally, in the same work, we discovered the mechanism of evolution that governs the emergence of SARS-CoV-2 variants  \cite{chen2020mutations}. Specifically,  we revealed that RBD mutations are governed by natural selection, when there were only 89 RBD mutations and their highest observed frequency was only about 50 globally (see Figure 3 of Ref. \cite{chen2020mutations}) in June 2020 \cite{chen2020mutations}. Stated differently, mutations that can increase virus infectivity will prevail. Based on 506,768 sequences isolated from patients, we demonstrated in  April 2021 that the 100 most observed RBD mutations out of 651 known RBD mutations all have their predicted mutation-induced binding free energy (BFE) changes above the average BFE change of -0.28kcal/mol \cite{wang2021vaccine}.  The odds for all these 100 most observed mutations to accidentally have the above-average  BFE changes is smaller than one chance in 1.27 nonillions, i.e., 1.27$\times10^{30}$. 
 
The objective of this review is to offer a general introduction of the mechanisms of SARS-CoV-2 evolution and to forecast emerging vaccine-breakthrough SARS-CoV-2 variants. We first discuss how mutations impact  SARS-CoV-2 infectivity. The molecular mechanism of viral infectivity and some of our work on mutation-induced infectivity changes are reviewed. Additionally, we discuss how the interplay of mechanisms at three scales, i.e., molecular, organism, and population scales, impact SARS-CoV-2 evolution. We demonstrate the co-existence of two complementary transmission pathways of viral evolution, i.e.,   infectivity and vaccine-resistant.  Finally, we review our studies on mutational impacts to SARS-CoV-2 antibodies and vaccines.

\section{Mutational impacts on SARS-CoV-2 infectivity}
\begin{figure}

	\centering
	\includegraphics[width=0.8\textwidth]{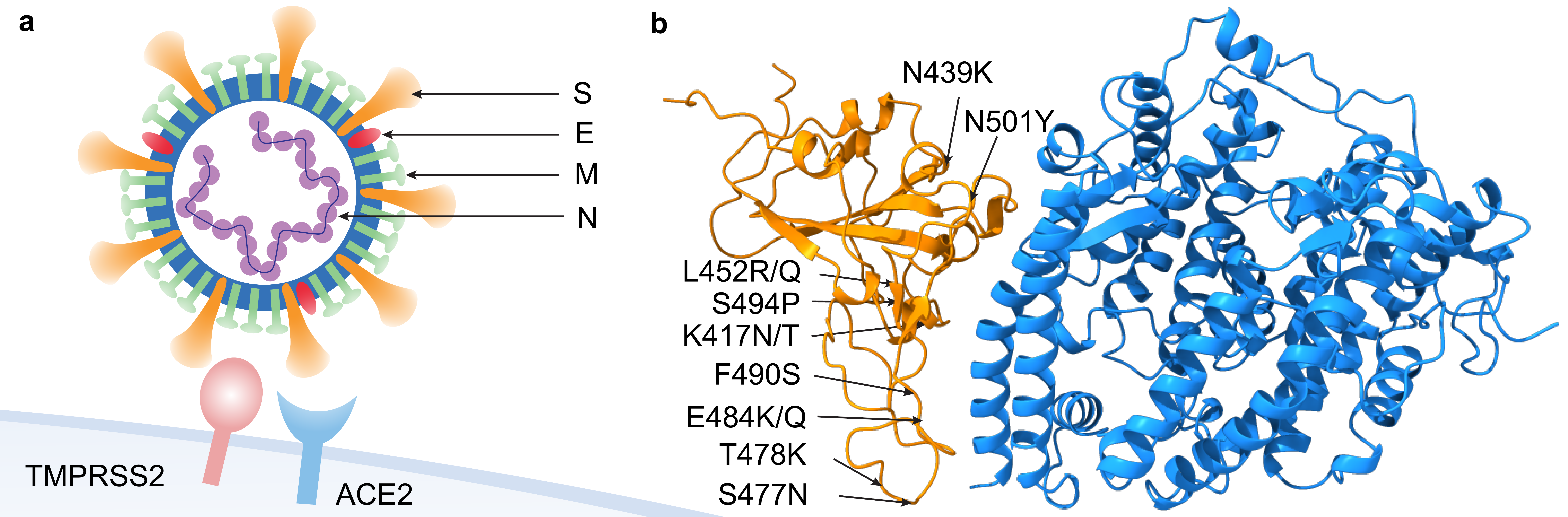}
    \caption{Illustration of SARS-CoV-2  and its interaction with host cell.  
		{\bf a} SARS-Cov-2 spike protein and ACE2 interaction. Here, viral spike (S), envelop (E), membrane (M), and nucleocapsid (N) proteins and host cell TMPRSS2 and ACE2  are labeled. 
		{\bf b} SARS-CoV-2 S protein RBD-ACE2 complex (PDB 6M0J) and important RBD mutations.  
		}
    \label{fig:6M0J_3D}
\end{figure}
 SARS-CoV-2 infectivity is a key factor in  COVID-19  prevention and control, and economic reopening. Viral infectivity can be referred to as the viral infection rate of a population, which depends on disease prevention measures, such as social distancing, use of masks, quarantine, contact tracing, etc. However,  viral intrinsic infectivity is a more fundamental concept and can be determined by experiments, i.e., counting the number of viruses in a specific volume over a unit of time by using either traditional or modern methods \cite{hoffmann2020sars}. The former includes plaque assay, focus forming assay, endpoint dilution assay, protein assay, hemagglutination assay, bicinchoninic acid assay, single radial immunodiffusion assay, and transmission electron microscopy. The latter has tunable resistive pulse sensing, flow cytometry, quantitative polymerase chain reaction, and enzyme-linked immunosorbent assay (ELISA)  \cite{glowacka2011evidence,hoffmann2020sars}. Traditional methods are generally slow and labor-intensive, while modern methods can significantly reduce viral quantification time. Among these methods, ELISA is based on protein-protein interactions (PPIs), such as antibody-antigen binding events, being counted by chromogenic or fluorescence reporters.  In the studies of SARS-CoV, epidemiological and biochemical studies show that the infectivity of different SARS-CoV strains in host cells is proportional to the  BFE between the  S protein  RBD    and the  ACE2  expressed by the host cell (see Figure \ref{fig:6M0J_3D}) 
 \cite{li2005bats,qu2005identification,song2005cross,hoffmann2020sars,walls2020structure}.  ACE2 is a single-pass transmembrane protein with its active domain exposed on the cell surface and is expressed in the lungs and many other tissues.  It is the main cell entry point for SARS-CoV and SARS-CoV-2, and some other coronaviruses. As shown in Figure \ref{fig:6M0J_3D}, SARS-CoV-2 cell entry is primed by TMPRSS2 (transmembrane serine protease 2)\cite{hoffmann2020sars}. 

 Both traditional and modern methods for viral infectivity measurement are expensive and time-consuming. For the rapidly spreading COVID-19 caused by the constantly mutating SARS-CoV-2, it is extremely challenging to experimentally determine the virus infectivity associated with all SARS-CoV-2 variants found around the world. Consequently, no such comprehensive experimental quantification program has been established for all RBD mutations so far. 

We established a computational platform to quantitatively predict RBD-ACE2 BFE changes induced by RBD mutations \cite{chen2020mutations}.   Our platform consists of the genotyping of SARS-CoV-2 genetic sequences isolated for  patients \cite{wang2020decoding0}, sequence alignment \cite{chen2020mutations}, the algebraic topology representation of PPIs \cite{cang2017topologynet,nguyen2020review}, the deep learning modeling of mutation-induced RBD-ACE2 BFE changes \cite{wang2020persistent}, and  the training using existing mutational datasets.  Based on genotyping, we have developed a real-time interactive \href{https://users.math.msu.edu/users/weig/SARS-CoV-2_Mutation_Tracker.html}{Mutation Tracker} (see Figure \ref{fig:mutationTracker}) to trace all SARS-CoV-2 mutations from complete genome sequences deposited at the GISIAD database (\url{https://www.gisaid.org/}). We unveiled that numerous mutations on commonly used SARS-CoV-2 diagnostic targets have caused false-negative test results \cite{wang2020mutations}, which was confirmed experimentally \cite{artesi2020recurrent}.  We decoded the ramification of mutations on SARS-CoV-2 vaccines, antibody therapies, and drugs \cite{wang2020decoding0}. We found that the top 8 mutations in the United States (US) belong to two groups. One group consisting of 5 concurrent mutations is prevailing,  while the other group with three concurrent mutations is gradually fading \cite{wang2021analysis}.

\section{Mechanisms of SARS-CoV-2 evolution}

Our approach to SARS-CoV-2 infectivity modeling is, by no means, inclusive. Many potential factors, such as the insertion of multi-basic residues at the S1/S2
subunit cleavage site \cite{hoffmann2020multibasic}, the high-frequency mutation D614G \cite{korber2020tracking}, temperature and pH conditions \cite{chan2020factors}, and numerous other hypotheses, have been extensively studied in the literature since the outbreak of the pandemic. However, none of these studies leads to the fundamental understanding of SARS-CoV-2 global transmission trajectory, infection pattern, and evolution mechanism.    

SARS-CoV-2 evolution is an intricate process determined by the interplay among molecular-scale mechanisms, organism scale mechanisms, and population-scale mechanisms. Mutagenesis is a  basic biological process at the molecular scale that changes the genetic information of organisms and thus, is a primary source for viral evolution. SARS-CoV-2 is a positive-strand RNA virus and belongs to the beta coronavirus genus with a genome size of 29.99 kb. Figure \ref{fig:mutationTracker} shows that SARS-CoV-2 has over 28,685 unique mutations. Therefore, on average, each SARS-CoV-2 nucleotide has one known mutation. At the molecular scale, mutations are the consequence of random genetic drifts during the replication, transcription errors due to limited polymerase fidelity,  ribosome fidelity associated translation errors, and various other errors caused by adversarial cellular environments generated by the host immune responses \cite{sanjuan2016mechanisms}. Additionally, at the organism scale,  host gene editing is a dominating mechanism for SARS-CoV-2 mutations  \cite{wang2020host}. However, unlike other single-stranded RNA viruses, such as flu virus and HIV, SARS-CoV-2  has a genetic proofreading mechanism achieved by an enzyme called non-structure protein 14 (NSP14) in synergy with  NSP12, i.e., RNA-dependent RNA polymerase \cite{gribble2021coronavirus}. Therefore, at the same condition, SARS-CoV-2 may not mutate as fast as flu viruses. Nonetheless, at the population scale, the evolution of SARS-CoV-2 is regulated by the mechanism of transmission. Viral infectivity and virulence are vital factors that determine viral transmission. On the other hand, the rapid, wide, and global transmission of SARS-CoV-2 fuels its widespread mutations.
 
 \begin{figure}
	\centering
	\includegraphics[width=0.6\textwidth]{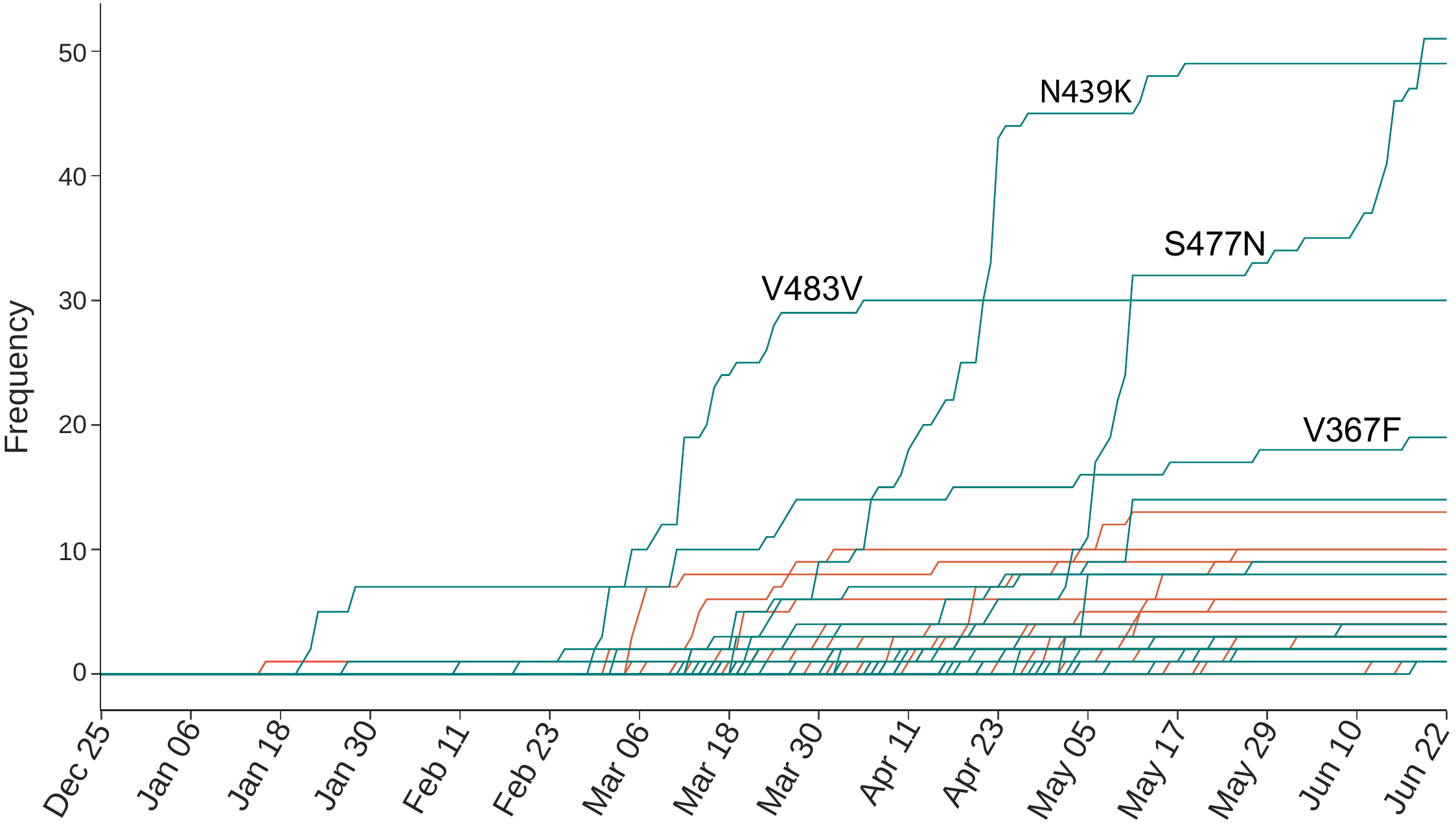}
    \caption{Reproduction of Figure 3 of Reference  \cite{chen2020mutations}. The time evolution of 89 SARS-CoV-2 S protein RBD mutations. The green lines represent the mutations that strengthen the infectivity of SARS-CoV-2, and the red lines represent the mutations that weaken the infectivity of SARS-CoV-2. Many mutations overlap their trajectories. Here, the collection date of each genome sequence deposited in GISAID was applied according to the information recorded in June 2020  \cite{chen2020mutations}.  
		}
    \label{fig:lineplot}
\end{figure}
 
In July 2020, the COVID-19 infection cases were about 12 million globally. We carried out the single nucleotide polymorphism (SNP) calling of over 15,000 SARS-CoV-2 genomes isolated from patients to identify 725  nondegenerated mutations on the SARS-CoV-2 S protein, including 89 mutations that occurred on the RBD, which are important to the binding of SARS-CoV-2 S protein and ACE2 (Our arXiv version was dated May 27, 2020). We found that most of these mutations have strengthened the RBD binding with ACE2 using our mathematical model based on a SARS-CoV-2 S protein RBD-human ACE2 complex structure from the Protein Data Bank (PDB ID 6M0J),  leading to more infectious SARS-CoV-2 strains. We stated that ``we hypothesize that natural selection favors those mutations that enhance the viral transmission and if our predictions are correct, the predicted infectivity strengthening mutations will outpace predicted infectivity weakening mutations over time.'' To our best knowledge, this work was the first of its kind that associates RBD mutations with COVID-19 population-level observations and natural selection mechanisms.  Figure \ref{fig:lineplot} recaps Figure 3 in our original paper \cite{chen2020mutations}. Since there were only 89 RBD mutations and their highest frequency was only about 50, statistics were not very convincing when we discovered natural selection as the SARS-CoV-2 mechanism of evolution. 
However, if our hypothesis is correct, our mechanism of SARS-CoV-2 evolution is viable, and our predictions of BFE changes induced by RBD mutations were sufficiently accurate, one would see the dramatic increase in frequencies of infectivity-strengthening RBD mutations globally as the pandemic unfolded in the following months.

\begin{figure}
	\centering
	\includegraphics[width=1\textwidth]{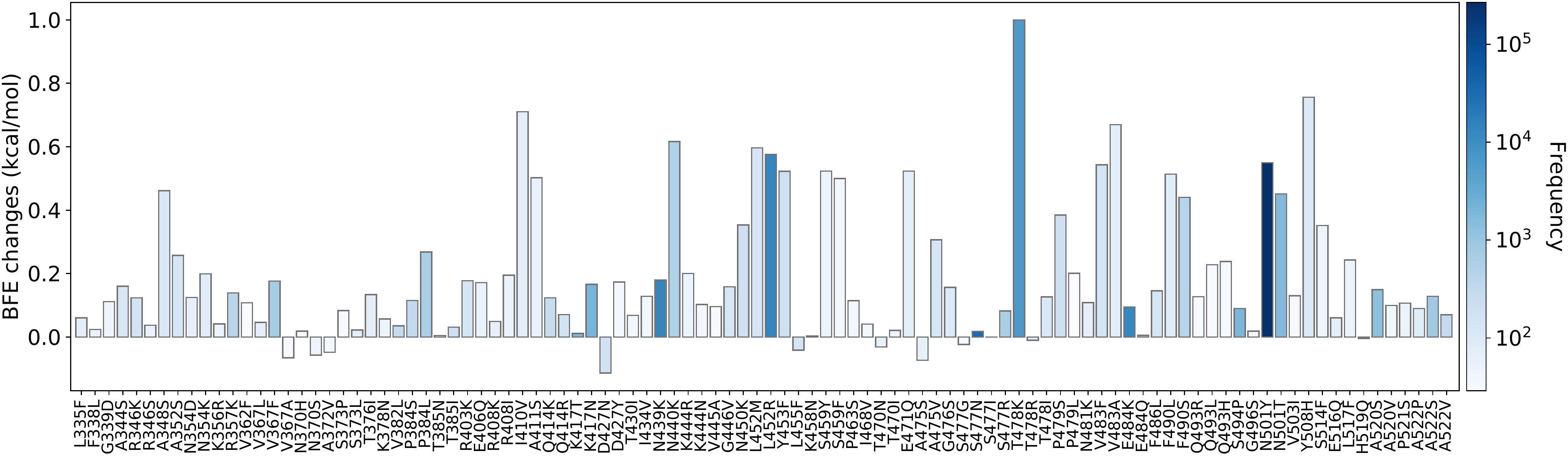}
    \caption{Reproduction of Figure 2 of Reference \cite{wang2021vaccine}. Illustration of SARS-CoV-2 mutation-induced BFE changes for the complexes of S protein and ACE2. Here, 100 most observed mutations out of 651 mutations on S protein RBD and their frequencies are illustrated as recorded in April 2021 \cite{wang2021vaccine}. The highest frequency was  168,801, while the lowest frequency was 28. Therefore, the frequencies of the rest of 551 mutations were lower than 28.  
		}
    \label{fig:top100s}
\end{figure}
 
 By April 2021,  the COVID-19 infection cases had increased to over 140 million. Most of the infections at that time were due to variants now known as Alpha, Beta, and Gamma. Their RBD (co-)mutations are [N501Y], [K417N, E484K, N501Y], and [K417T, E484K, N501Y], respectively. The mutation at residue 501 has significantly strengthened the viral infectivity as predicted by us in May 2020 \cite{chen2020mutations}. In the next few months, the Delta variant, involving RBD mutation L452R and T478K,  had rapidly fueled a new wave of COVID-19 infections around the world. The mutation at residue 452, as predicted by us in May 2020 \cite{chen2020mutations}, increases the RBD-ACE2 BFE by 0.55 kcal/mol, making the virus more infectious \cite{wang2021vaccine}. Additionally, our prediction indicates that T478K increases the RBD-ACE2 BFE by 1.00 kcal/mol \cite{wang2021vaccine}. The combined effect of L452R and T478K makes the Delta variant the most infectious variant among all variants formally named by the World Health Organization (WHO).   
  
Since early 2021, there has been a dramatic increase in the effort to sequence SARS-CoV-2 genomes. By April 2021, our  Mutation Tracker (\url{https://users.math.msu.edu/users/weig/SARS-CoV-2_Mutation_Tracker.html}) had recorded 651 RBD mutations and their highest frequency was 168,801 based on 506,768 SARS-CoV-2 genome sequences isolated from patients \cite{wang2021vaccine}. Additionally, some deep mutational data on RBD and ACE2 became available in the literature around early 2021 \cite{starr2020deep, linsky2020novo}, which significantly improved the predictive accuracy of our deep learning model \cite{chen2021prediction}.  Therefore, we were in a much better position in April 2021 to revisit the evolution mechanism discovered in July 2020. Figure  \ref{fig:top100s} recaps the Figure 2 in our paper published in May 2021 \cite{wang2021vaccine}.  The 100 most observed RBD mutations were shown to have BFE above the average values of $-0.28$kcal/mol. The odd for this to happen as an accident is one chance in $2^{100}$. Therefore, our hypothesis that natural selection is a principle mechanism for SARS-CoV-2 evolution was proven beyond all doubts in April 2021.

\section{Mutational impacts on SARS-CoV-2 antibodies and vaccines}

One of the most important issues concerning SARS-CoV-2 variants is how mutations will escape existing vaccines and disrupt known mAbs. A more intriguing question is whether vaccination will alter the mechanisms of SARS-CoV-2 transmission and induce vaccine-resistant variants. It is imperative to understand how the emerging variants impact vaccines. The host immune response to a common vaccine depends on race, gender, age, health condition, personal genetic background, etc. Therefore, antibodies induced by vaccination differ from person to person. 
For a given vaccine, it is impossible to estimate how many different antibodies will be generated in a population. However, the statistics of various antibodies can be used to study the efficacy of vaccination in a population. We collected seven antibody-RBD complex  structures extracted from COVID-19 patients  in March 2020 \cite{chen2021review} and extended the number of antibody-RBD complex structures to 56 in March 2021 \cite{chen2021prediction}. Currently, we have a library of collections of 130 antibody-RBD complex structures \cite{wang2021emerging} and use these complexes as a statistical ensemble to understand how RBD mutations will impact antibodies and vaccines.

 \begin{figure}
	\centering
	\includegraphics[width=0.35\textwidth]{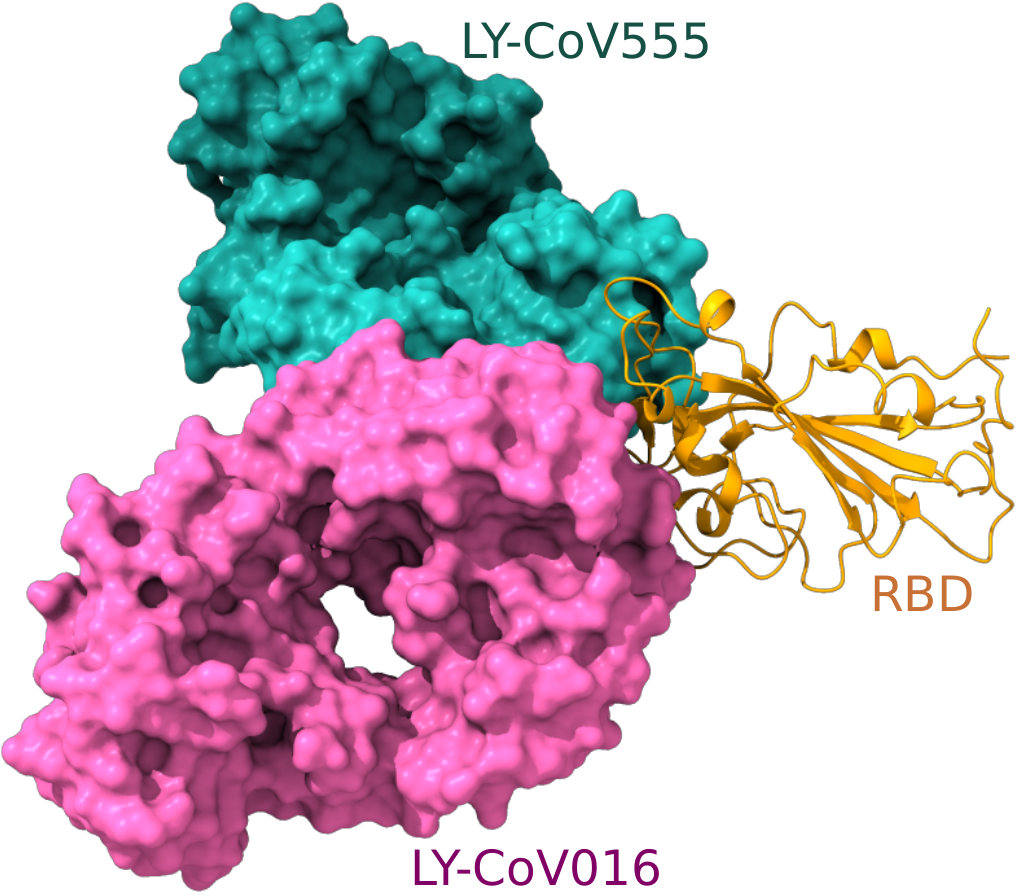}
    \caption{Reproduction of the right chart of Figure 11 of Reference \cite{chen2021revealing}. The S protein RBD is in a binding complex with Eli Lilly antibodies LY-CoV555 (PDB 7KMG) and LY-CoV016 (PDB 7C01).
		}
    \label{fig:LY-CoV}
\end{figure}

 \begin{figure}
	\centering
	\includegraphics[width=1\textwidth]{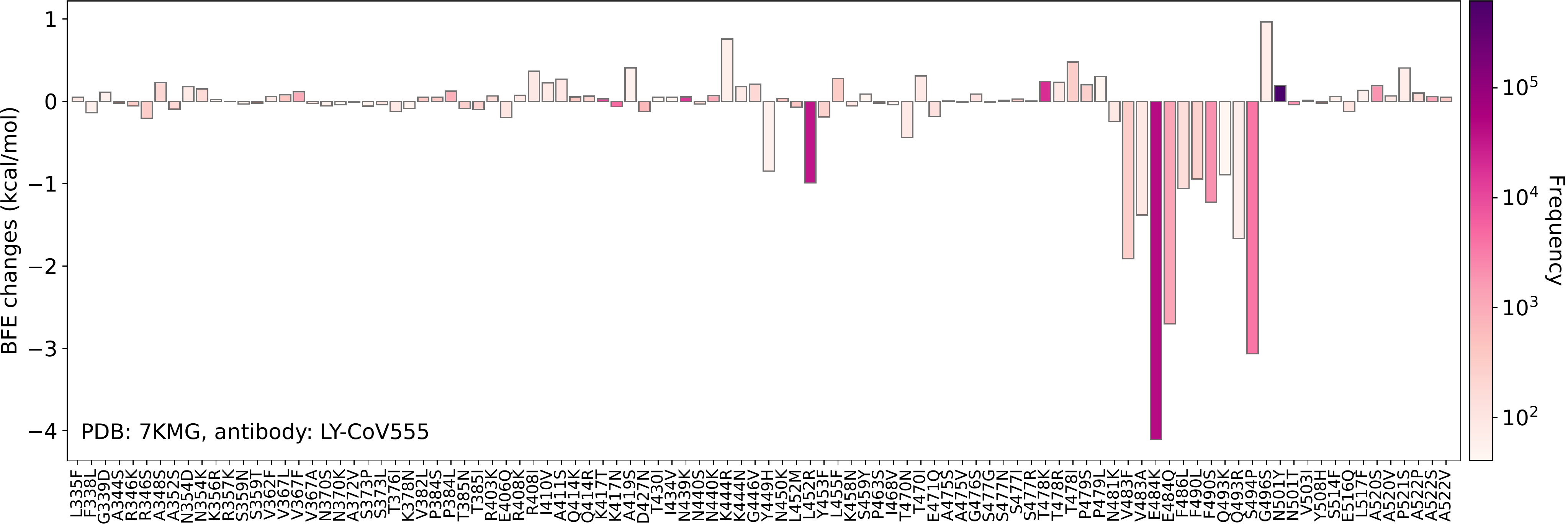}
    \caption{Reproduction of Figure 12 of Reference \cite{chen2021revealing}. 
	 The BFE changes of S protein and antibody LY-CoV555 (PDB 7KMG) induced by the 100 most observed RBD mutations are shown.  Here, mutations L452R, V483F/A, E484K/Q, F486L, F490L/S, Q493K/R, and S494P were shown to potentially disrupt the binding of antibodies and S protein RBD.	}
    \label{fig:7KMG_bar}
\end{figure}

\begin{figure}
	\centering
	\includegraphics[width=1\textwidth]{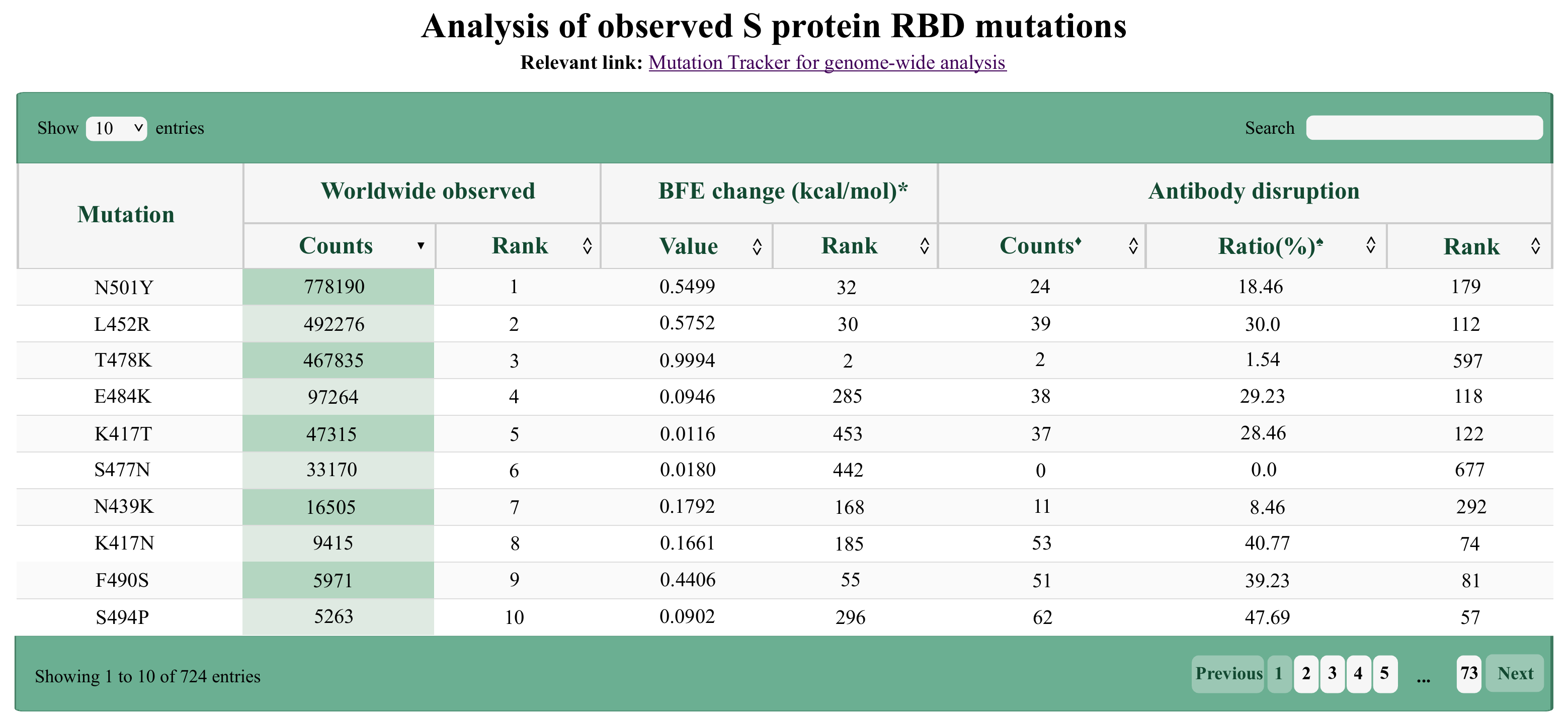}
    \caption{Illustration of the analysis of SARS-CoV-2 mutations given by   Mutation Analyzer. Interactive version is available at web site:  \url{https://weilab.math.msu.edu/MutationAnalyzer/}. 
		}
    \label{fig:MutationAnalyzer}
\end{figure}
Most recently, we have studied how RBD mutations and various variants will impact mAbs that have obtained the emergency use authorization (EUA) from the FDA, i.e.,  those from Regeneron and Eli Lilly and mAbs in clinical trials, including those from Celltrion and Rockefeller University  \cite{chen2021revealing}.  Figure \ref{fig:LY-CoV} shows the binding complex of RBD and Eli Lilly mAbs. The impacts of the 100 most observed RBD mutations to the binding complex of RBD and antibody LY-CoV555 are demonstrated in Figure \ref{fig:7KMG_bar}. We noted that L452R, V483F/A, E484K/Q, F486L, F490L/S, Q493K/R, and S494P could disrupt Eli Lilly mAbs. Indeed, Eli Lilly mAbs were taken off the market in March due to Beta [K417N, E484K, N501Y] and Gamma [K417T, E484K, N501Y] variants. However, Eli Lilly mAbs were allowed to return to the market in September because they are quite effective on the Delta variant [452R, T478K]. Our predictions in  Figure \ref{fig:7KMG_bar} provide a good explanation. An extensive comparison of experimental data and our predictions of mutational impacts to mAbs was given in Ref. \cite{chen2021revealing}. A comprehensive analysis of the 75 most observed RBD mutational disruptions of 130 existing antibodies was presented in our recent work \cite{wang2021emerging}. Our analysis of all RBD mutations is available at an interactive website, Mutation Analyzer (\url{https://weilab.math.msu.edu/MutationAnalyzer/}) as shown in Figure \ref{fig:MutationAnalyzer}.

\section{Vaccination impacts on SARS-CoV-2 evolution }

\begin{table}[h!]
    \centering
    \setlength\tabcolsep{1pt}
	\captionsetup{margin=0.1cm}
	\caption{RBD  mutation analysis of WHO-named SARS-CoV-2 variants and emerging variants (EV). 
	Antibody disruption (AD) count and  BFE change (kcal/mol)  are given.}
    \label{tab:Variants}
    {
    \begin{tabular}{l|l|c|c }
    \toprule
    Variants & ~RBD (co-)mutation(s) & AD count  &  BFE change\\
    \midrule
    Alpha&[N501Y]&24&0.55\\
    Beta&[K417N, E484K, N501Y]&90&0.81\\
    Gamma&[K417T, E484K, N501Y]&81&0.66\\
    Delta&[L452R, T478K]&40&1.58\\
    Epsilon&[L452R]&39&0.58\\
    Zeta&[E484K]&38&0.10\\
    Eta&[E484K]&38&0.10\\
    Theta&[E484K, N501Y]&60&0.65\\
    Iota&[E484K]&38&0.10\\
    Kappa&[L452R, E484Q]&52&0.58\\
    Lambda&[L452Q, F490S]&59&1.42\\
    Mu&[R346K, E484K, N501Y]&64&0.77\\   
		\midrule
		EV& [F490S, N501Y] &72 & 0.99\\
		EV& [S494P, N501Y] & 73& 0.64\\
		EV& [K417N, L452R, T478K] &82 &1.74\\
		EV& [L452R, T478K, N501Y] &59 &2.13\\
	  EV& [A411S, L452R, T478K] & 49&2.08\\
		EV& [L452R, T478K, E484K, N501Y] &75 &2.22\\
		EV& [P384L, K417N, E484K, N501Y] & 101&1.08\\
    \bottomrule 
    \end{tabular}
    }
\end{table}

 \begin{figure}
	\centering
	\includegraphics[width=1\textwidth]{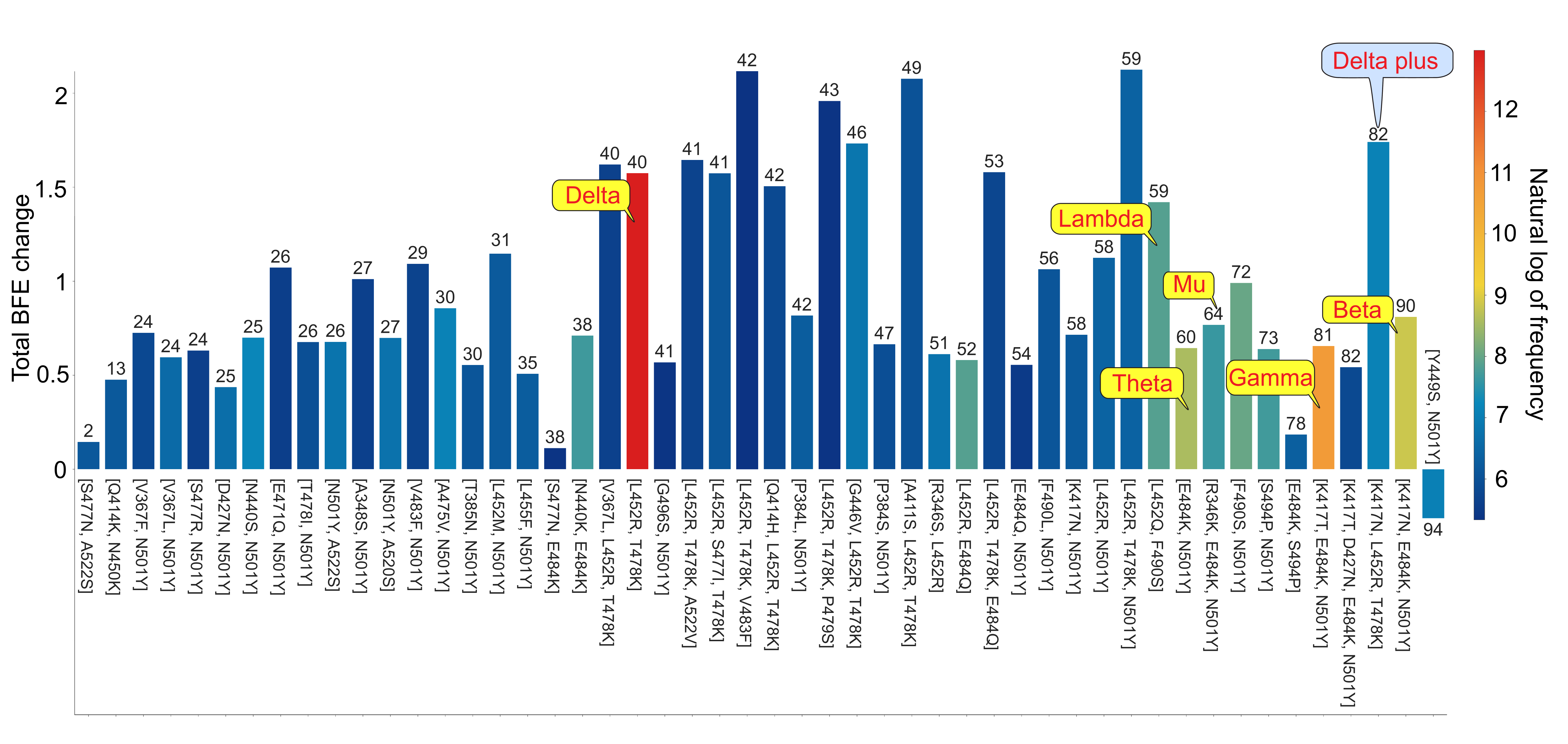}
    \caption{Analysis of RBD co-mutations. Here, the $x$-axis lists  RBD co-mutations and the $y$-axis represents the predicted total BFE change of each set of RBD co-mutations. The color of each bar represents the natural log of frequency for each set of RBD co-mutations. The number on the top of each bar is the AI-predicted number of antibody and RBD complexes that may be significantly disrupted by the set of RBD co-mutations with BFE change smaller than $-0.3$ kcal/mol.  \cite{wang2021vaccine}.
	  	}
    \label{fig:comutations}
\end{figure}

Currently,  the WHO has named variants from Alpha to Mu.  All of these variants 
strengthen the binding between RBD and ACE2 as shown in Table \ref{tab:Variants}.   Judged by RBD (co-)mutations, the Delta variant is the most infectious one. 
The Lambda variant is as infectious as the Delta variant, followed by the Beta variant. Zeta, Eta, and Iota are the least infectious in the list. In terms of their ability to disrupt antibodies,  Beta and Gamma variants are among the top, followed by new variant Mu. The Alpha variant has the lowest ability to escape vaccines.   Therefore, Beta and Gamma variants are more likely to break through current vaccines. Note that the variant Lambda is not only as infectious as the Delta, but also capable of disrupting many antibodies,  and thus, is potentially more dangerous than all other WHO-named variants.   
A list of vaccine-escape (vaccine-breakthrough) mutations was tabulated in our early work\cite{wang2021vaccine}, including S494P, Q493L, K417N, F490S, F486L, R403K, E484K, L452R, K417T, F490L, E484Q, and A475S.

An important trend is that many variants involve two or three RBD co-mutations. Indeed, the Delta variant consists of two infectivity strengthening mutations. Table \ref{tab:Variants} also lists a few RBD co-mutations that do not have corresponding WHO-named variants at present. Many of these co-mutations are about multiples of ten times more infectious than the original SARS-CoV-2.   
Figure \ref{fig:comutations} provides a summary of the 50 most observed RBD co-mutation sets presented in Figure 2 of our recent work \cite{wang2021emerging}. We noticed  that only five out of 50 sets of RBD co-mutations do not involve any of residues 452 and 501. Future variants that combine RBD co-mutations at 452 and 501 will be extremely contagious.  Delta variant has the highest global frequency, following by Gamma, Beta, etc.  All the most observed RBD co-mutations, except one on the right, have positive BFE changes and strengthen the viral infectivity. Therefore, these co-mutation sets also support our hypothesis that the evolution of SARS-CoV-2 is governed by infectivity-based natural selection.

Worth noting that the co-mutation set [Y449S, N501Y] has a negative BFE change of $-0.26$ kcal/mol, highlighting its infectivity weakening characteristic. We postulated that an alternative transmission pathway had led to the occurrence of co-mutation set [Y449S, N501Y] \cite{wang2021emerging}. Indeed, this co-mutation set has the highest antibody disruption count of 94 among the 50 most observed RBD co-mutation sets, showing its exceptionally strong ability to break through current vaccines. Our further analysis indicated that this co-mutation set has kicked off around May 2021 in countries having a high level of vaccination, including  Denmark, the  United Kingdom, etc. Therefore, we conclude that vaccine-breakthrough is a new transmission pathway for SARS-CoV-2 evolution and it will become a dominating pathway for emerging SARS-CoV-2 variants when populations fully carry SARS-CoV-2 antibodies either through vaccination or via infection.  

Figure \ref{fig:comutations} reveals a  general trend that co-mutation sets on the right side, with relatively high antibody disruption counts have relatively high frequencies. Many of these co-mutation sets involve vaccine-escape mutations predicted in our early work: S494P, Q493L, K417N, F490S, F486L, R403K, E484K, L452R, K417T, F490L, E484Q, and A475S \cite{wang2021vaccine}. It becomes clear that emerging SARS-CoV-2 variants will tend to combine high infectivity with strong vaccine-breakthrough capability. Based on this observation,  we anticipate that co-mutation sets [A411S, L452R, T478K], [L452R, T478K, N501Y] and [K417N, L452R, T478K] are on the path to becoming emerging SARS-CoV-2 variants. Additionally,  although their frequencies are relatively low at present, co-mutation sets  [L452R, T478K, E484K, N501Y] and [P384L, K417N, E484K, N501Y] have very high BFE changes (i.e., 2.22 and 1.07 kcal/mol,  respectively) and very high antibody disruption counts (i.e., 75 and 101, respectively)\cite{wang2021emerging}.   We foresaw that these co-mutation sets would also pose a severe threat. 

\section{Methods and validations}
 
\begin{figure}
	\centering
	\includegraphics[width=0.7\textwidth]{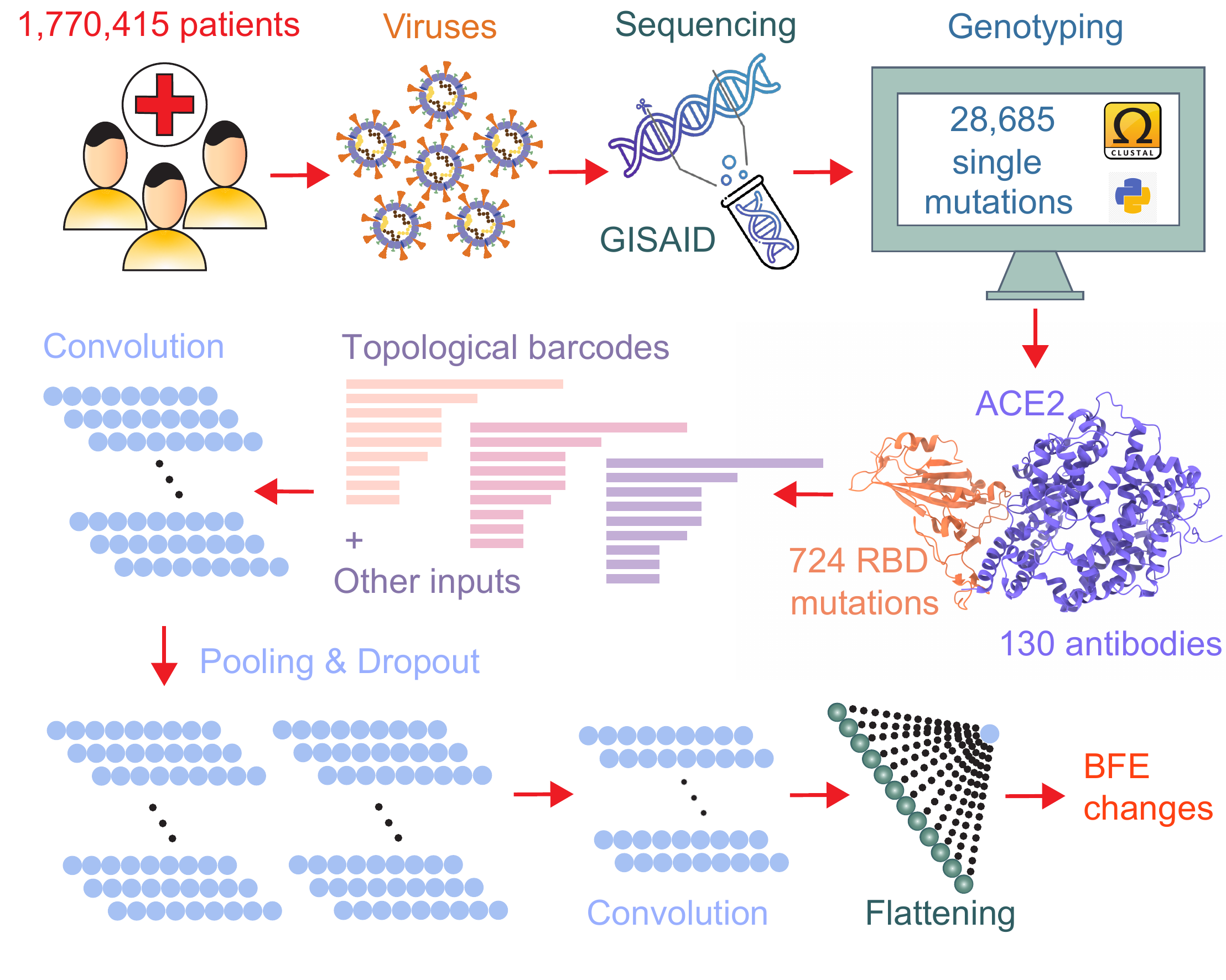}
    \caption{  Illustration of genome sequence data pre-processing and BFE change predictions \cite{wang2021emerging}.  	}
    \label{fig:flowchart}
\end{figure}

Figure \ref{fig:flowchart} provides a bird's eye view of the data collection and computational platform introduced in our predictions and comparison with some experimental data. First, viral genomes isolated COVID-19 patient samples are sequenced and deposited in the GISIAD database. We download 1,770,415 complete SARS-CoV-2 genome sequences with high coverage and exact collection date from the GISAID database \cite{shu2017gisaid}.  For each sequence, we carried out SNP calling to obtained mutation information for the whole genome based on a reference SARS-CoV-2 genome \cite{wu2020new}. The Cluster Omega (\url{https://www.ebi.ac.uk/Tools/msa/clustalo/}) is used for multiple sequence alignment. We share our genotyping results with the research community using our interactive \href{https://users.math.msu.edu/users/weig/SARS-CoV-2_Mutation_Tracker.html}{Mutation Tracker}.  

We employ the structure of the RBD-ACE2 complex (6M0J) to predict RBD mutation-induced BFE changes \cite{chen2020mutations,chen2021prediction}, which provides mutation-induced infectivity change estimations. In our recent work, we have collected a library of 130 antibody-RBD complexes, which are used to predict mutational impacts to existing antibodies \cite{chen2021review,chen2021prediction,wang2021vaccine,wang2021emerging,chen2021revealing}. These antibodies were extracted from COVOD-19 patients and their detailed information is available elsewhere \cite{wang2021emerging}. 

 In  our deep learning models, persistent homology \cite{carlsson2009topology,edelsbrunner2008persistent,meng2020weighted} is used   to represent RBD-ACE2 and RBD-antibody interactions. Element-specific persistent homology offers a controllable simplification of biomolecular complexity and retains crucial physical and biological information of PPIs \cite{cang2017topologynet,nguyen2020review}. Additional features were also generated to complement topological ones \cite{wang2020topology,chen2020mutations}. In our early work, a deep learning algorithm called NetTree, which combines convolutional neural networks (CNNs) and gradient boosting trees (GBTs), was constructed to tackle the challenge of scarce data in our early predictions \cite{wang2020topology,chen2020mutations}. Our early model was trained with a large dataset of 8,338 PPI entries from the SKEMPI 2.0 dataset \cite{jankauskaite2019skempi,wang2020topology}. In our recent work, our deep learning models were trained with additional deep mutational datasets associated with SARS-CoV-2 RBD \cite{starr2020deep, linsky2020novo}, and ACE2\cite{procko2020sequence}. Additionally, we made use of deep mutational data on both proteins for the binding complex of RBD and protein  CTC-445.2 \cite{linsky2020novo}. Our results are   available at our interactive website \href{https://weilab.math.msu.edu/MutationAnalyzer/}{Mutation Analyzer} \cite{chen2021revealing}.

\begin{figure}
	\centering
	\includegraphics[width=1\textwidth]{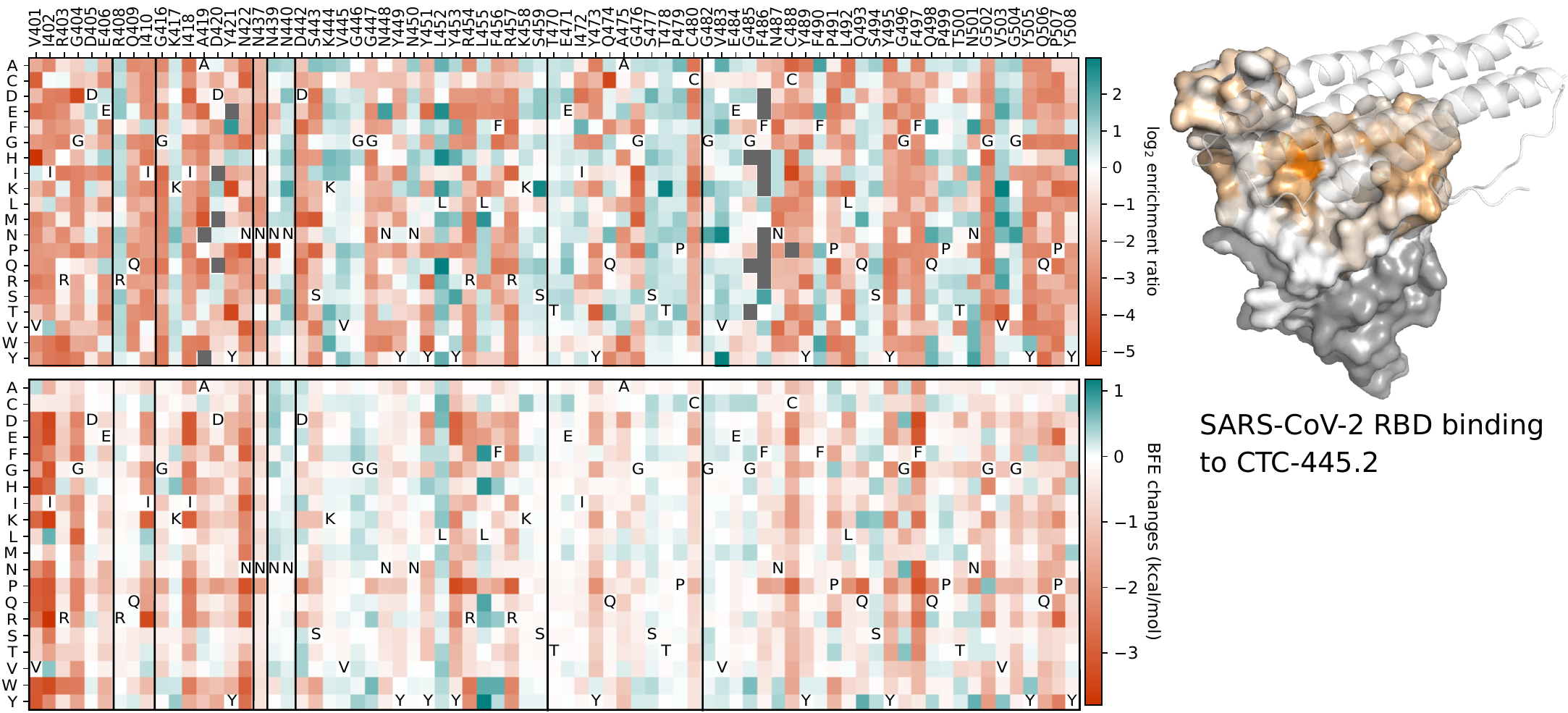}
    \caption{ Reproduction of Figure 2 of reference \cite{chen2021prediction} for  a comparison between experimental deep mutation enrichment data and TopNetTree predictions for the SARS-CoV-2 S protein RBD and CTC-445.2 complex (7KL9 \cite{linsky2020novo}). Top left: deep mutational scanning heatmap showing the average effect on the enrichment for single-site mutants of the RBD when assayed by yeast display for binding to CTC-445.2.89. Top right: the RBD colored by average enrichment at each residue position bound to CTC-445.2. Bottom: machine learning predicted BFE changes for the CTC-445.2 and S protein complex induced by single-site mutations on the RBD.}
    \label{fig:7KL9_RBD_combine}
\end{figure}

The extensive validations of our deep learning predictions were presented in our recent papers \cite{chen2021prediction,chen2021revealing, wang2021emerging}.  Figure \ref{fig:7KL9_RBD_combine} shows our prediction of RBD mutation-induced BFE chances for a complex of protein CTC-445.2 and spike protein RBD. Our predictions were systematically compared with experimental deep mutational enrichment data  \cite{chen2021prediction,linsky2020novo}.  Additionally, a similar comparison between experimental RBD deep mutation enrichment data and predicted BFE changes for SARS-CoV-2 RBD binding to ACE2 (6M0J) was also shown a high-level correlation in our other work \cite{chen2021revealing,linsky2020novo}. Moreover, extensive validations were carried out with experimental data of the impacts of emerging variants on antibodies in clinical trials \cite{chen2021revealing}. Further, our predicted BFE changes induced by RBD mutations L452R and N501Y for ACE2 and RBD complex have a highly similar trend with experimental data in relative luciferase units \cite{chen2021revealing,deng2021transmission}.  Finally, a comparison of experimental CT-P59 IC$_{50}$ fold change (reduction) \cite{lee2021therapeutic} and predicted BFE changes induced by mutations L452R and T478K showed a high correlation \cite{wang2021emerging}. 

\section{Conclusion}
 
Understanding the mechanisms of evolution and transmission of severe acute respiratory syndrome coronavirus 2 (SARS-CoV-2) is a challenging task and a prerequisite to predict future variants. In this review, we demonstrate that the SARS-CoV-2 mechanism of evolution and transmission was discovered one year ago. In July 2020, we unveiled that mutations on the SARS-CoV-2 spike (S) protein receptor-binding domain (RBD) strengthened viral infectivity, which serves a unique viral transmission pathway \cite{chen2020mutations}. We revealed that infectivity-based natural selection governs SARS-CoV-2 evolution. In April 2021, we had proven beyond all doubt that infectivity-based natural selection was the sole mechanism for SARS-CoV-2 evolution and transmission. However, we show that the mechanism of SARS-CoV-2 evolution also depends on an alternative transmission pathway. We identify that an infectivity-weakening RBD co-mutation set [Y449S, N501Y] stands out among heavily vaccinated populations because it has an exceptionally strong ability to break through existing vaccines, signaling a new mechanism of SARS-CoV-2 evolution and transmission. Our studies integrate genotyping of over 1.77 million SARS-CoV-2 genomes isolated from patients, a library collection of 130 antibodies, tens of thousands of SARS-CoV-2 related deep mutational experimental data, algebraic topology, and deep learning.        

In the same paper published one year ago \cite{chen2020mutations}, we also predicted mutations at RBD residues 452 and 501 would ``have high chances to mutate into significantly more infectious COVID-19 strains.'' It is now known that RBD mutations at these sites are essential for prevailing variants  Alpha, Beta, Gamma, Delta, Epsilon, Theta, Kappa, Lambda, and Mu.  We anticipate that RBD mutations at 452 and 501 will be persistently vital to the next generation of new variants. The combination of RBD co-mutations at 452 and 501 will lead to new variants that are about multiples of ten times more infectious than the original SARS-CoV-2 and have a much higher ability to escape current vaccines.   

Currently, SARS-CoV-2 is evolving mainly through the infectivity-based transmission pathway in unvaccinated and less infected populations. In future, when most of world's populations carry SARS-CoV-2 antibodies either through vaccination or infection, the antibody-disruption-based transmission pathway of viral evolution will become more important.   These two complementary transmission pathways of viral evolution will govern the emergency of emerging SARS-CoV-2 variants, which will extend our battle with COVID-19 for a long time. We forecast that a few co-mutation sets, including [A411S, L452R, T478K], [L452R, T478K, N501Y], [K417N, L452R, T478K],   [L452R, T478K, E484K, N501Y],  and  [P384L, K417N, E484K, N501Y], have a great potential to grow into unprecedentedly dangerous new SARS-CoV-2 variants.  We call for the development of the next generation of vaccines to prevent these new mutations.    

\section*{Data and codes availability}

Mutational data for complete SARS-CoV-2 genomes are available for download at our interactive web page Mutation Tracker (\url{https://users.math.msu.edu/users/weig/SARS-CoV-2_Mutation_Tracker.html}), which is updated regularly. 

Observed frequencies (counts) of all spike protein receptor-binding domain (RBD) mutations, RBD mutation-induced binding free energy (BFE) changes of the angiotensin-converting enzyme 2 (ACE2) and RBD complex,  and counts of RBD mutation-induced disruptions of RBD and antibody complexes are given by our interactive web page Mutation Analyzer (\url{https://weilab.math.msu.edu/MutationAnalyzer/}), which is updated regularly.

As a review, this work does not contain new codes. However, related codes are available in cited references.

\section*{Acknowledgments}
This work was supported in part by NIH grant  GM126189, NSF grants DMS-2052983,  DMS-1761320, and IIS-1900473,  NASA grant 80NSSC21M0023,  Michigan Economic Development Corporation, MSU Foundation,  Bristol-Myers Squibb 65109, and Pfizer.


 
 \bibliographystyle{abbrv}
 \bibliography{refs}

\end{document}